\documentclass[aps,twocolumn]{revtex4}
\usepackage{adjustbox}
\usepackage{epsfig}
\usepackage{graphicx}
\usepackage{amsmath,amssymb,amsfonts}
\usepackage{array}
\usepackage{url}
\usepackage{hyperref}
\usepackage{multirow}
\usepackage{float}
\usepackage{lineno}
\usepackage{xspace}
\usepackage[usenames,dvipsnames]{color}
\newcommand{\sqsn}{\mbox{$\sqrt{s_{_{NN}}}$}\xspace}
\newcommand{\pt}{$p_{\rm{T}}$}
\newcommand{\nch}{$N_{\rm ch}$}
\newcommand{\bef}{\begin{figure}}
\newcommand{\eef}{\end{figure}}
\newcommand{\bc}{\begin{center}}
\newcommand{\ec}{\end{center}}

\newcommand{\be}{\begin{equation}}
\newcommand{\ee}{\end{equation}}
\newcommand{\bea}{\begin{eqnarray}}
\newcommand{\eea}{\end{eqnarray}}

\begin{document}

\title{Dynamics of particle production in Pb--Pb collisions at $\sqrt{s_{NN}}$ = 2.76 TeV using PYTHIA8 Angantyr model}
\author{Ravindra Singh}
\author{Yoshini Bailung}
\author{Ankhi Roy}
\affiliation{Discipline of Physics, Indian Institute of Technology Indore, Simrol, Indore 453552, India}

\begin{abstract}

We study the dynamics of identified, strange, and multi-strange particle production in Pb--Pb collisions at \sqsn = 2.76 TeV using the recently developed Angantyr model, incorporated within PYTHIA8. We show the interplay between multi-parton interactions (MPI) and color reconnection (CR) on the experimentally measured quantities. The charged-particle multiplicity (\nch) and mean transverse momentum ($\langle$\pt$\rangle$) distributions are well explained by PYTHIA8 Angantyr with proper tuning, as presented in this paper. Predictions of \pt \ spectra, $\langle$\pt$\rangle$, \pt \ integrated yields of the identified, strange and multi-strange particles are studied. To provide insight into the collective nature of the produced particles, we look into the ratio of particle yields to pions and kaons. PYTHIA8 Angantyr with CR and MPI mimic signs of collectivity and is possibly one of the suitable candidates to study ultra-relativistic heavy-ion collisions.

\end{abstract}
\date{\today}
\maketitle

\section{Introduction}
\label{intro}

The study of how matter behaves at extreme temperature and energy densities is vital to understand the nature of phase transition in Quantum Chromodynamics (QCD)\citep{Gao:2020qsj,Tawfik:2016cot,Satz:2013xja}. Experiments like Relativistic Heavy Ion Collider (RHIC) at BNL, USA, and Large Hadron Collider (LHC) at CERN, Geneva, Switzerland, are of prime importance to unravel the properties of the hot QCD matter produced\citep{Busza:2018rrf,Teaney:2000cw,Cassing:1999es,PHOBOS:2004zne}. A de-confined state of quarks and gluons, also known as Quark-Gluon Plasma (QGP), is believed to be created for a very short lifetime in these heavy-ion collisions experiments. In this QGP phase, the relevant degrees of freedom are quarks and gluons rather than mesons and baryons, confined to color-neutral states~\cite{Becattini:2014rea,Busza:2018rrf,Aoki:2006we}. 

An initial and important observable is the multiplicity distribution ($\rm d\it{N}_{\rm ch}/d\eta$) of the charged particles, which is essential to extract the properties of produced particles and their interactions~\cite{PHOBOS:2004zne}. Such distributions in a particular pseudorapidity range were measured in the CERN proton anti-proton ($\rm p\overline{p}$) collider experiments in 1980's~\citep{UA5:1988gup,UA1:1989bou,UA5:1986yef,UA1:1982fux}. These measurements provide information on the energy density and centrality of the colliding system. The centrality is directly related to the initial overlap region geometry, which corrrespond to the number of participating nucleons and binary collisions\citep{Miller:2007ri}. 

For the final-state charged particles, the rapidity ($\rm{y}$) or pseudorapidity ($\eta$) and transverse momentum (\pt) spectra are known to reflect the degrees of longitudinal extension and transverse excitation of the interacting system, respectively\citep{PHENIX:2004vcz,Braun-Munzinger:2003pwq}. Distributions in low \pt \ ranges let us inspect the transverse excitation and soft processes, whereas higher \pt \ correspond to hard scattering processes. In low energy collisions, one can neglect hard processes, while most of the contribution comes from soft processes. At high center of mass energies, hard processes have finite contribution albeit, the soft processes are predominant~\cite{Liu:2013uba}. From the final state charged particle \pt \ spectra, one can extract information about the thermal nature of the interacting system~\citep{Cleymans:2012ya,Cleymans:2011in}, and can comment on the formation and characteristic properties of the formed matter. According to the Maxwell-Boltzmann distribution law, the \pt \ spectra are related to the temperature of the system formed in these collisions. The \nch \ of charged particles formed in ultra-relativistic collisions also depends on the system's temperature and density. Since most of the final state charged particles are part of a locally thermalized medium, the mean transverse momentum $\langle$\pt$\rangle$ distribution vs. \nch \  is expected to be more or less flat in heavy-ion systems like Pb--Pb at high \nch . The contributions at lower \nch \ are mostly from the peripheral collisions where a QGP is less likely to be produced.

The ratios of yields of identified hadrons are important to understand to the mechanism of hadron production. The ratio of proton to pion ($p/\pi$) and kaon to pion ($K/\pi$) characterize the relative baryon and meson production, respectively. Additionally, $K/\pi$, $\Lambda/\pi$, $\Sigma^{0}/\pi$, $\Xi^{0}/\pi$, and $\Omega/\pi$ ratios represent the strangeness production at higher multiplicities, indicating a universal underlying dynamics in hadron production for different quark-gluon final states. Strangeness enhancement is proposed as a signature of QGP formation in heavy-ion collisions~\citep{Koch:1986ud,Soff:1999et,Capella:1994pr} because of the faster equilibration of strangeness production processes in a QGP than any other process in a hadron gas~\cite{Cuautle:2016huw,Torrieri:2009mq}. This is observed to be more prominent for multi-strange hadrons~\cite{ALICE:2016fzo}. The production mechanism of strange hadrons provides a way to investigate the properties of the hot QCD matter.

Another essential medium characteristic is anisotropic flow, considered the proof of collective behavior of partons and hadrons\citep{Huovinen:2001cy,ALICE:2010suc,STAR:2004jwm}. In a heavy-ion collision, the hydrodynamic expansion is a consequence of the transverse pressure gradient. This transverse flow shifts the produced particles to higher momenta, and due to higher gain in momenta of heavy particles from flow velocity, the increment is more for heavier particles. This effect is seen commonly for heavy-ion systems and even high multiplicity p--p and p--Pb collisions~\cite{ALICE:2016fzo}.

In this work, we have explored all the above-mentioned aspects of particle production in Pb--Pb collisions at \sqsn = 2.76 TeV using the Angantyr model, which is the heavy-ion extension of the PYTHIA8, extensively used for pp collisions. The paper is organized as follows. In section~\ref{evt_gen}, the details of event generation and analysis methodology are discussed. section~\ref{result} presents the results and discussions. Finally, in section~\ref{sum}, we summarize our findings of this investigation.

\section{Event generation and Analysis methodology}
\label{evt_gen}
PYTHIA is an event generator that is extensively and successfully used for the study of proton-proton and proton-lepton collisions. Recent advancement in PYTHIA8 enables the study of heavy nuclei collisions, namely proton-nuclei (pA) and  nuclei-nuclei (AA). In this work, PYTHIA8 event generator is used to simulate ultra-relativistic Pb--Pb collisions with Angantyr~\cite{Bierlich:2018xfw}. PYTHIA8 natively does not support heavy-ion systems; however, the Angantyr model combines several nucleon-nucleon collisions into one heavy-ion collision. It is a combination of many-body physics (theoretical) models suitable for producing hard and soft interactions, initial and final-state parton showers, particle fragmentations, multi-partonic interactions, color reconnection mechanisms, and decay topologies~\cite{Sjostrand:2006za}. We use 8.235 version of PYTHIA8~\cite{Sjostrand:2008za}, which includes multi-parton interactions (MPI). MPI is vital to expostulate the underlying events, multiplicity distributions, and charmonia production. In general, an event generator at high colliding energies produces around four to ten partonic interactions, which depend on the overlapping region of colliding particles~\cite{Weber:2018ddv}. The perturbative scattering processes are implemented by Initial State Radiation (ISR) and Final State Radiation (FSR)\citep{BaBar:2015onb,Martinez:2014cfa}.\\

Hadronization in PYTHIA8 is done using the Lund string fragmentation model. In this model, the probability of creation of hadrons from an initial state of partons is described by the Lund area law~\cite{Andersson:2001yu}. The beam remnants and the produced partons are interconnected via color fluxtubes or strings storing potential energy. New quark-antiquark pairs are formed as strings break when the partons move away from each other. Until the strings reduce to small pieces, this process continues. These small string pieces are recognized as shell hadrons.  In this scheme, strings are reconnected between the partons in such a way that strings length decreases; particle production decreases and consequently multiplicity reduces. In literature, flow-like feature in pp collisions is well mimicked by Color Reconnection (CR)~\cite{OrtizVelasquez:2013ofg} tune of PYTHIA8.\\

In the current version of Angantyr included in PYTHIA8, hadronization is done using the string fragmentation model; a thermalized medium of partons (QGP) is not a part of this model~\cite{Bierlich:2018xfw}.  In Angantyr, positions and the number of the interacting nucleons and binary nucleon-nucleon collisions are performed by Glauber model-based eikonal approximation in impact-parameter space. Furthermore, Gribov's corrections are implemented in order to include diffractive excitation, which appears due to the fluctuations in the nucleon substructure. Angantyr is the first model in which individual fluctuations for both projectile and target are included. It is required in order to extrapolate into AA systems~\cite{Bierlich:2018xfw}. The contribution to the final state from each participating nucleon is induced from the Fritiof model with the concept of wounded nucleons (diffractive and non-diffractive). The hard partonic sub-collisions, normalized by nucleon-nucleon sub-collisions, play a crucial role at high energies. The model treats the projectile and target nucleons via two interaction scenarios. In one case, the interactions between the species are considered as pp like non-diffractive (ND) processes. This is entirely driven by PYTHIA8. However, in the second scenario, a wounded projectile nucleon can have ND interactions with many target nucleons, which are termed secondary ND (SD) collisions. Subevents are generated solely through PYTHIA8, where these SD collisions are put into play as modified SD processes~\cite{Behera:2019cds}. Depending on the interaction probability, interactions between wounded nucleons are classified as elastic, non-diffractive, secondary non-diffractive, single diffractive, and double-diffractive.\\ \\
We have generated around 2 million events in Pb--Pb at \sqsn = 2.76 TeV. The inelastic, non-difractive component of the total cross-section for all soft QCD processes is used with the switch \texttt{SoftQCD:all = on} with MPI based scheme of color reconnection (\texttt{ColorReconnection:mode(0)}). The classes based on charged particle multiplicities (\nch) have been chosen in the acceptance of TPC detector with pseudorapidity range of ($-0.8<\eta<0.8$)~\cite{ALICE:2010suc} to match with experimental conditions and centrality classes in ALICE at the LHC. The events generated using these cuts are divided into nine multiplicity classes, each class containing 10\% of total events except first two classes which contain 5\% of total events as used in \cite{ALICE:2010suc}. The \nch \ classes corresponding to different centralities are tabulated in TABLE~\ref{tab:TPC}. Reconstruction of heavy strange particles are done via specific decay channels or their weak decay topologies, namely $ \Xi^{\mp} \xrightarrow{} \pi^{\mp} + \Lambda$, $ \Omega^{\mp} \xrightarrow{} K^{\mp} + \Lambda $, ${\Phi \xrightarrow{} \mu^{-} + \mu^{+} }$, ${\Sigma^{0} \xrightarrow{} \Lambda + \gamma }$ , and  ${\Lambda \xrightarrow{} \pi^{-} + p }$  and for the anti-particles we used corresponding charge conjugate decay channels. We use -0.8  $< \eta < 0.8 $ and -0.9  $< y < 0.9 $ kinematics range for reconstruction these particles to account for ALICE experimental acceptances.

\begin{table}[h]
\caption{Centrality classes and the corresponding charged particle multiplicities.}
\centering 
\scalebox{1.1}
{
\begin{tabular}{|c|c|c|} 
\hline 
 S.No. & Centrality (\%)& \nch \\
\hline 
I &0-5 & 2314-3050\\
\hline
II &5-10 &1947-2314\\
\hline
III &10-20 & 1387-1947\\
\hline
IV &20-30 &967-1387\\
\hline
V &30-40 & 644-967\\
\hline
VI &40-50 & 399-644\\
\hline
VII &50-60 & 224-399\\
\hline
VIII &60-70 & 108-224\\
\hline
IX &70-80 & 43-108\\
\hline
\end{tabular}
}
\label{tab:TPC}
\end{table}

\begin{table}[h]
\caption{Mean and RMS of charge particle multiplicity in different PYTHIA8 tunes.\\}
\centering 
\scalebox{1.05}
{
\begin{tabular}{|c|c|c|c|c|} 
\hline 
    & MPI + CR & CR off & MPI off & MPI off + CR off \\
\hline 
Mean & 704.1 & 882.8 & 276.6 & 276.6\\
\hline
RMS & 759.5 & 961.2 & 274.3 & 274.3 \\
\hline
\end{tabular}
}
\label{tab:meanrms}
\end{table}

\section{Results and Discussion} 
\label{result}

\begin{figure*}[ht!]
\includegraphics[scale=0.43]{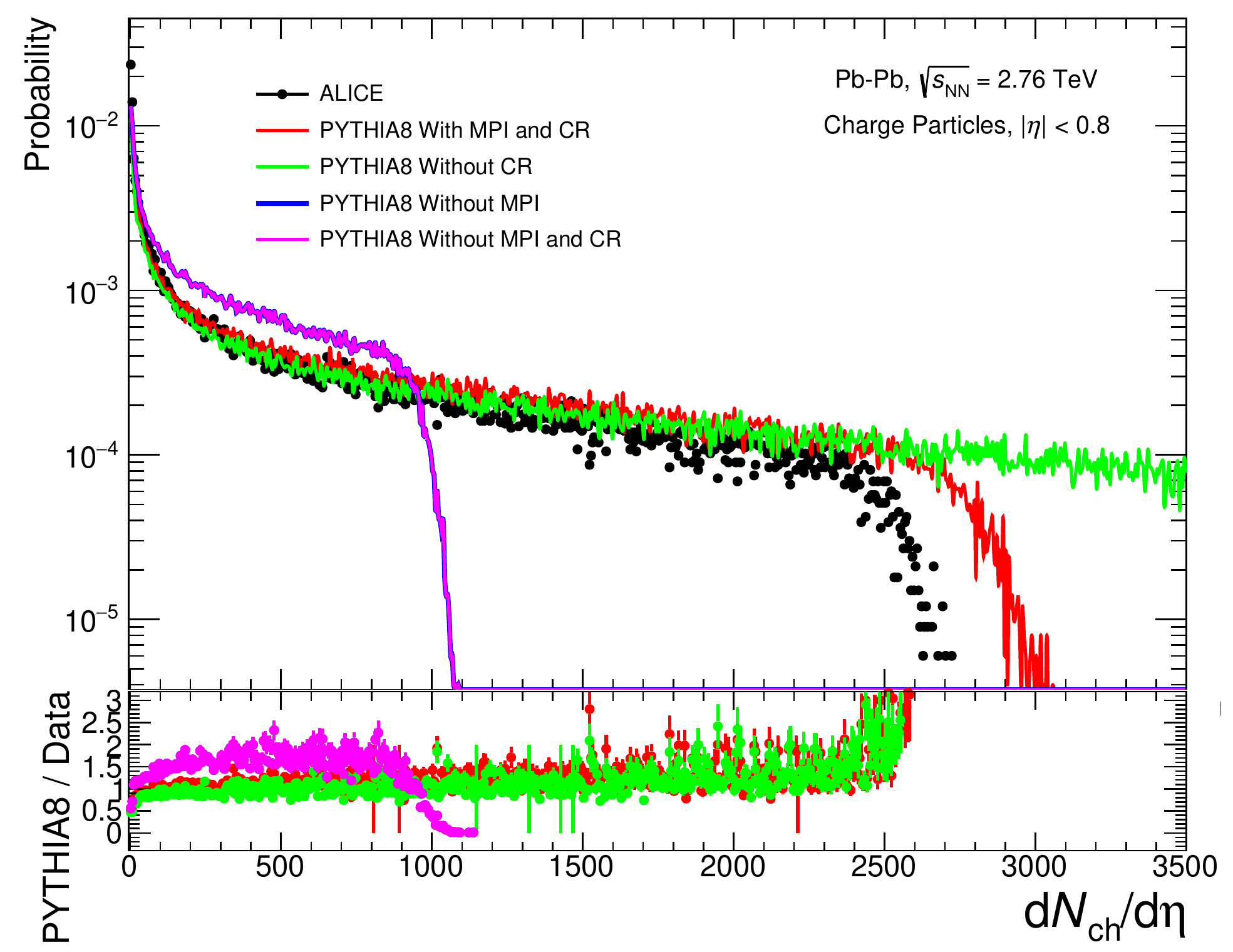}
\includegraphics[scale=0.43]{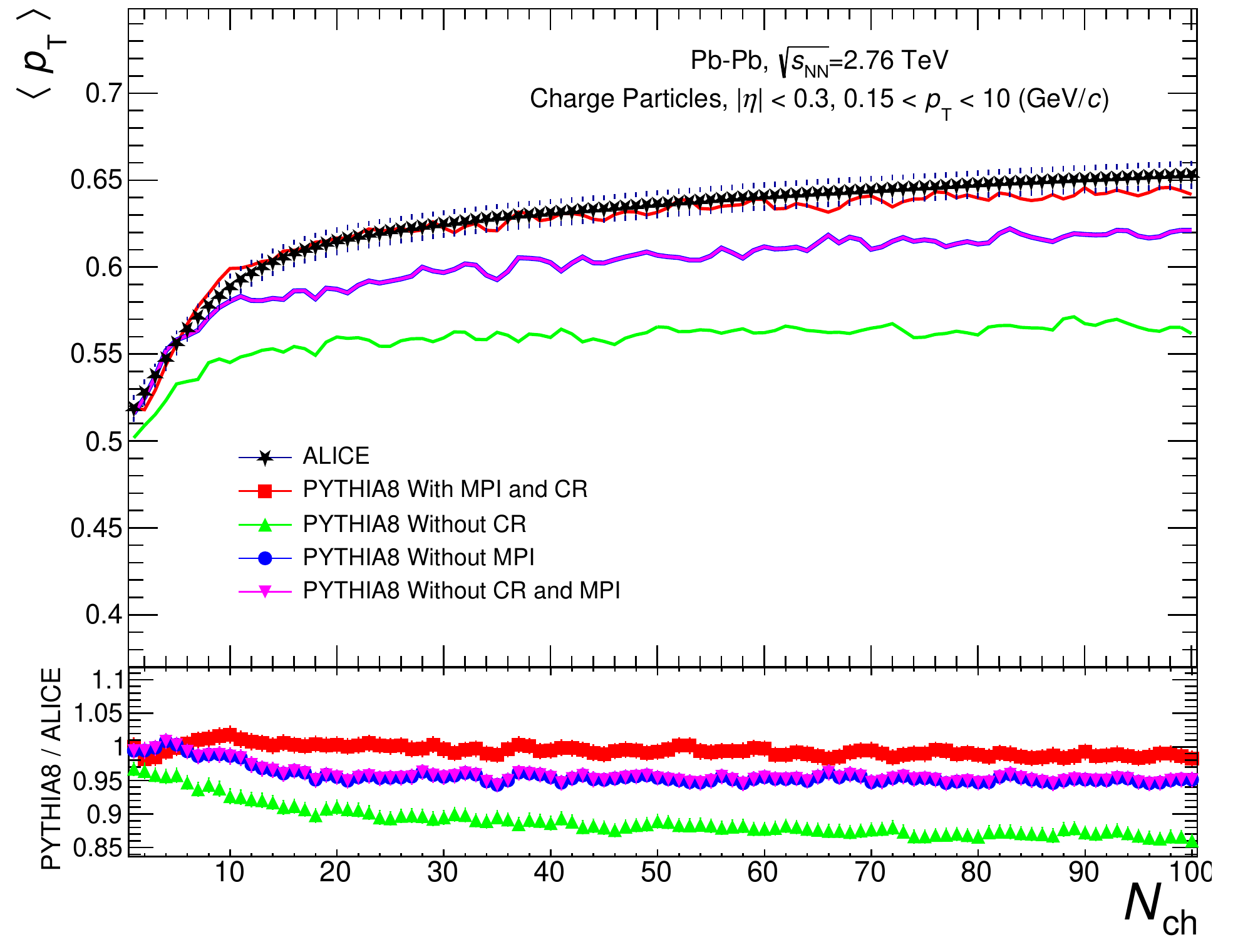}
\caption{(Color Online) (Left) Multiplicity distribution of charged particles from PYTHIA8 Angantyr with different tunes and ALICE data. (Right) $\langle$\pt$\rangle$ distribution vs. charged-particle multiplicity in different PYTHIA8 tunes and ALICE data. The lower panels show the ratio of PYTHIA Angantyr predictions over Data for the different configurations considered.}
 \label{fig1}  
\end{figure*}

The charged-particle multiplicity distributions for different PYTHIA8 tunes within $|\eta|<0.8$) is  shown in the FIG.~\ref{fig1}. To see the effect of different PYTHIA tunes, we consider the following configurations: MPI with/without CR, No MPI, and both MPI and CR off. It is observed that MPI+CR is compatible with ALICE data. MPI without CR overestimates, whereas the tune without MPI is seen to underestimate our results. We also observe that there is no effect of CR if MPI is off. The particle production increases with MPI, due to inter-partonic interactions; on the other hand, while turning CR off, particle production increases. In the color reconnection (CR) scheme, the string lengths are reduced; in consequence, when CR is kept on, particle production lessens~\cite{OrtizVelasquez:2013ofg}. Turning MPI off removes the strings between the partons. As a result, we do not observe any effect of CR. To quantify the effect of MPI and CR for the different PYTHIA tunes, the mean and RMS of the multiplicity distributions are measured and reported in TABLE~\ref{tab:meanrms}. The mean for MPI+CR is around 2.5 times larger without MPI and around 3.2 times larger without CR compared without MPI. A similar comparison can be made for the RMS values between these settings. For MPI+CR turned off, we report similar values for mean and RMS, which confirms our statement made earlier.

By observing different tunes in FIG.~\ref{fig1} (Left), we can conclude that MPI with CR is a favorable setting to describe ALICE data~\cite{ALICE:2010suc}. We also observed similar results after comparing $\langle$\pt$\rangle$ distribution as a function of charged-particle multiplicity using simulated PYTHIA8 Angantyr and experimental data as shown in FIG.~\ref{fig1} (Right). Distributions obtained from PYTHIA8 are scaled with a constant (1.138) factor for better visualization and to compare the slope of different distributions with data~\cite{ALICE:2013rdo}. The $\langle$\pt$\rangle$ distribution with MPI and CR describes the data very well, even without hydrodynamics. $\langle$\pt$\rangle$ with MPI off (or MPI and CR off) describe data below $ N_{\rm ch} = 10$ very well but deviates at higher values, becoming almost flat at high multiplicities. This is probably due to the large production of low multiplicity events when MPI is kept off. A similar trend is seen for CR turned off; however, the ratio of $\langle$\pt$\rangle$ over data decreases as we go to higher values in multiplicity, as reconnection occurs in such a way that the strings between partons are as small as possible. This attribute is credited to CR, where a correlation between \nch \ and $\langle$\pt$\rangle$ can be seen~\cite{ALICE:2012aqc}. Preceeding hadronization, strings fuse to form high \pt \ hadrons.  With CR off, fewer strings fuse to form hadrons during hard scatterings, explaining the increment of $\langle$\pt$\rangle$ at higher multiplicities. 

\begin{figure*}[ht!]
\includegraphics[scale=0.42]{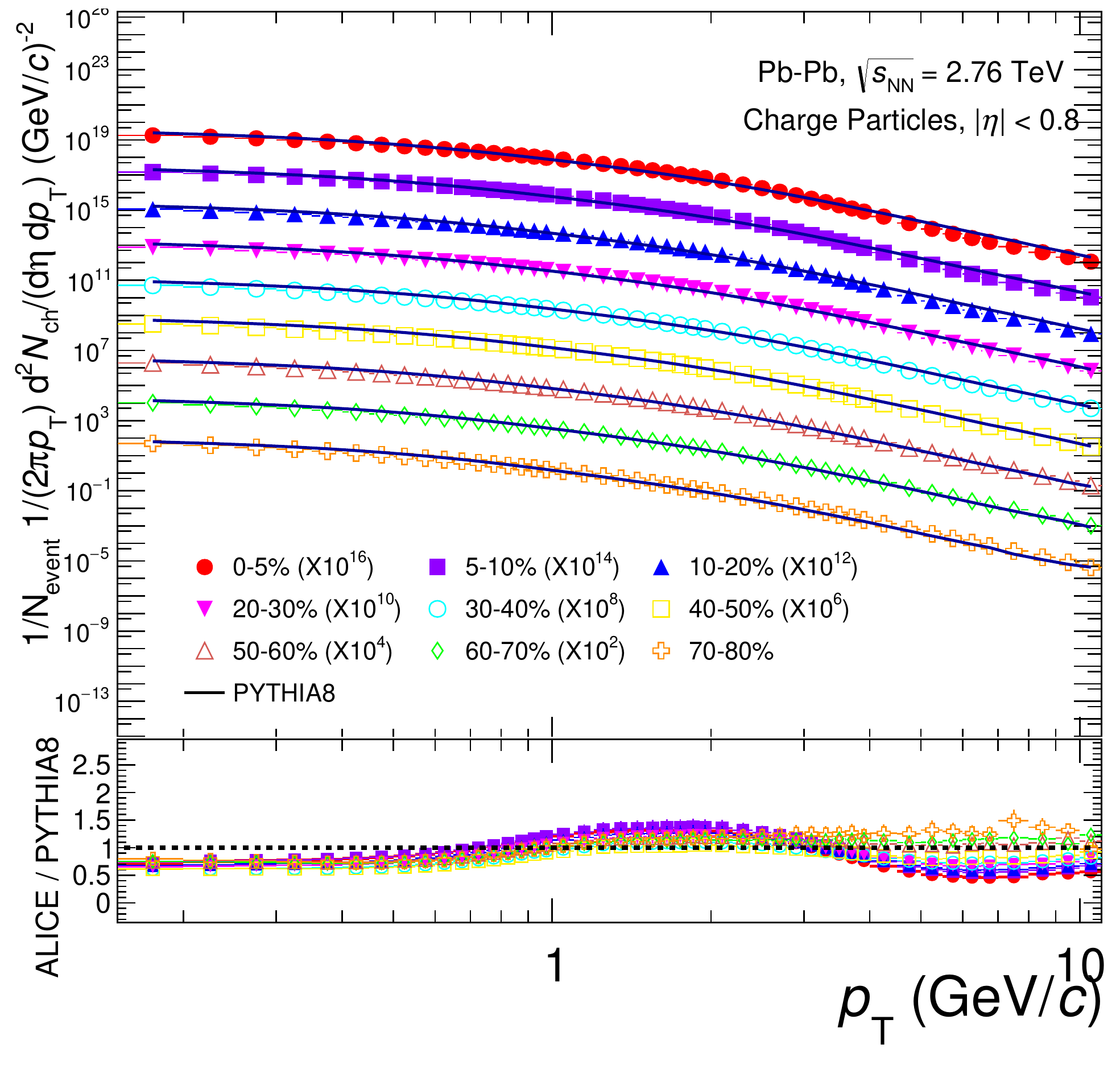}
\caption{(Color Online) Charged-particle \pt \ spectra in nine centrality classes described in TABLE~\ref{tab:TPC} from PYTHIA Angantyr and ALICE data. The lower panel represents the deviation of PYTHIA Angantyr predictions from data.}
\label{fig2}  
\end{figure*}

To further check the compatibility of simulated data, we compare the \pt \ spectra of final state charged-particles with ALICE measurements in different centrality classes within experimental kinematic selections, which is shown in FIG.~\ref{fig2}. 
The \pt \ spectra of each centrality class are scaled to the slope with ALICE measurements for clearer visualization and comparison. From the lower panel of FIG.~\ref{fig2}, it is observed that the experimental to simulated data is comparable within statistical uncertainties.

With the assurance of the quality of the simulated data discussed above, we now move on to study the transverse momentum spectra of identified particles, \pt \ integrated yield of identified and strange particles, and particle ratios with PYTHIA8.

\subsection{Transverse momentum spectra of identified particles}

\begin{figure*}[ht!]
\includegraphics[scale=0.365]{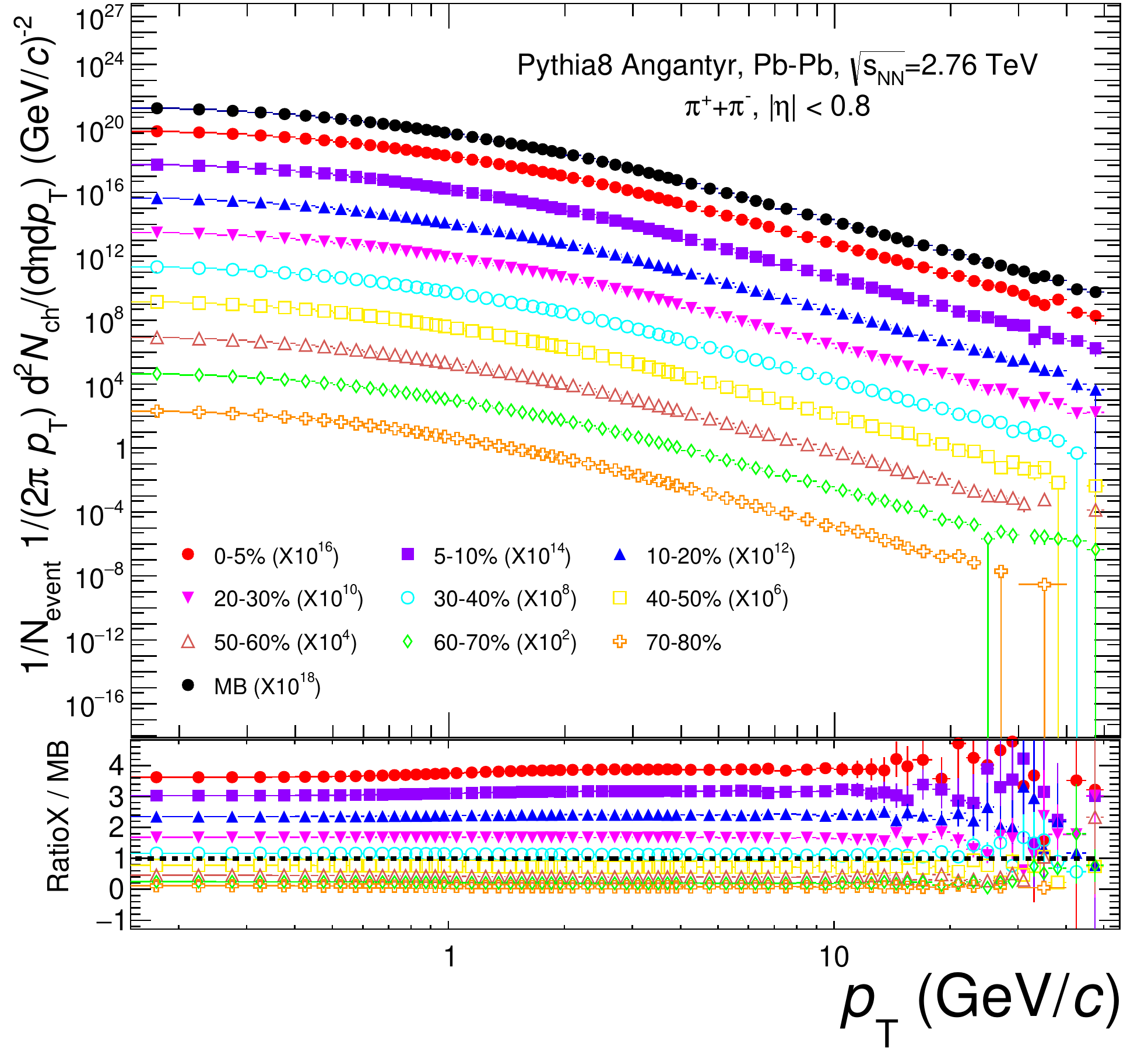}
\includegraphics[scale=0.365]{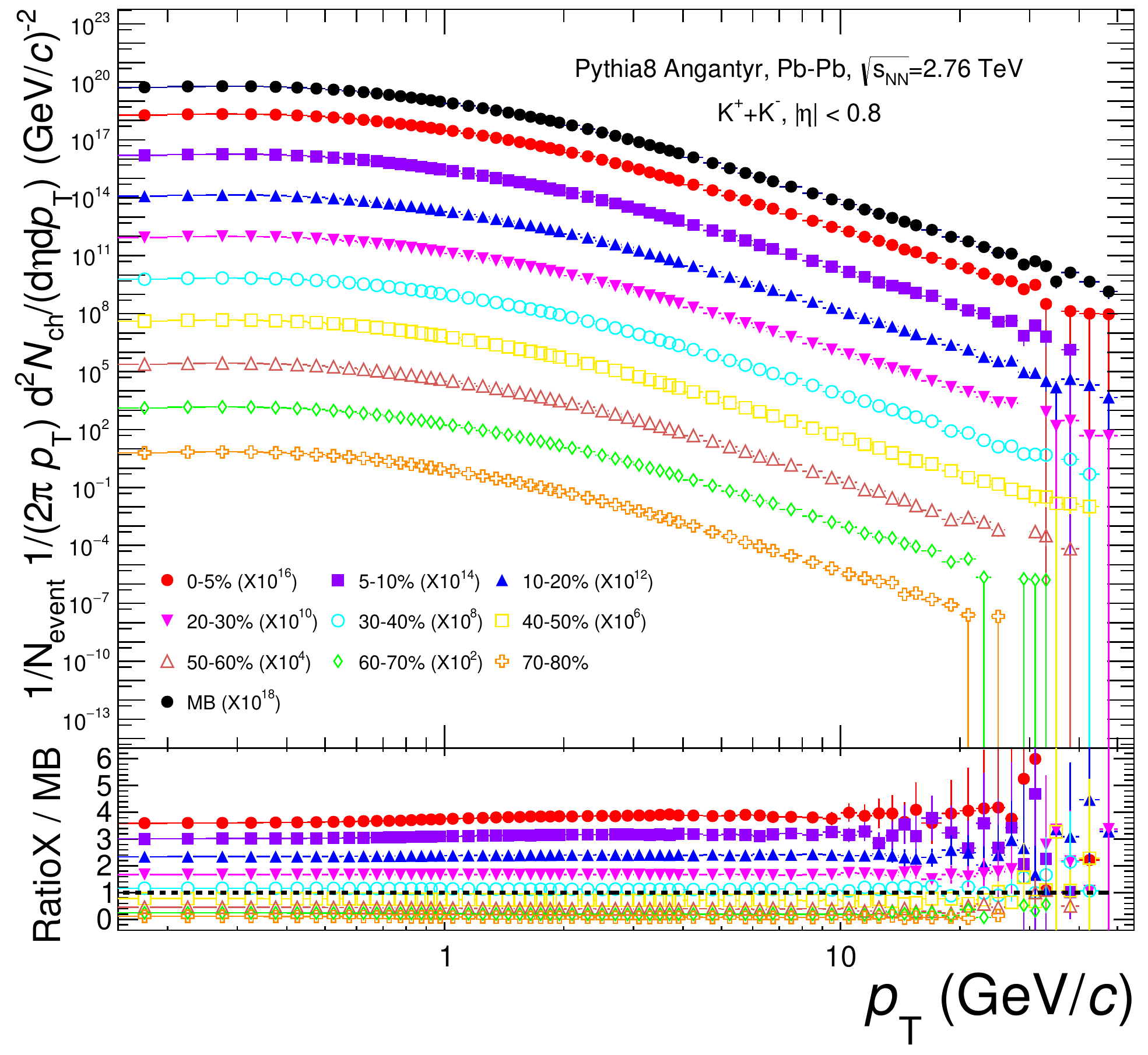}
\includegraphics[scale=0.365]{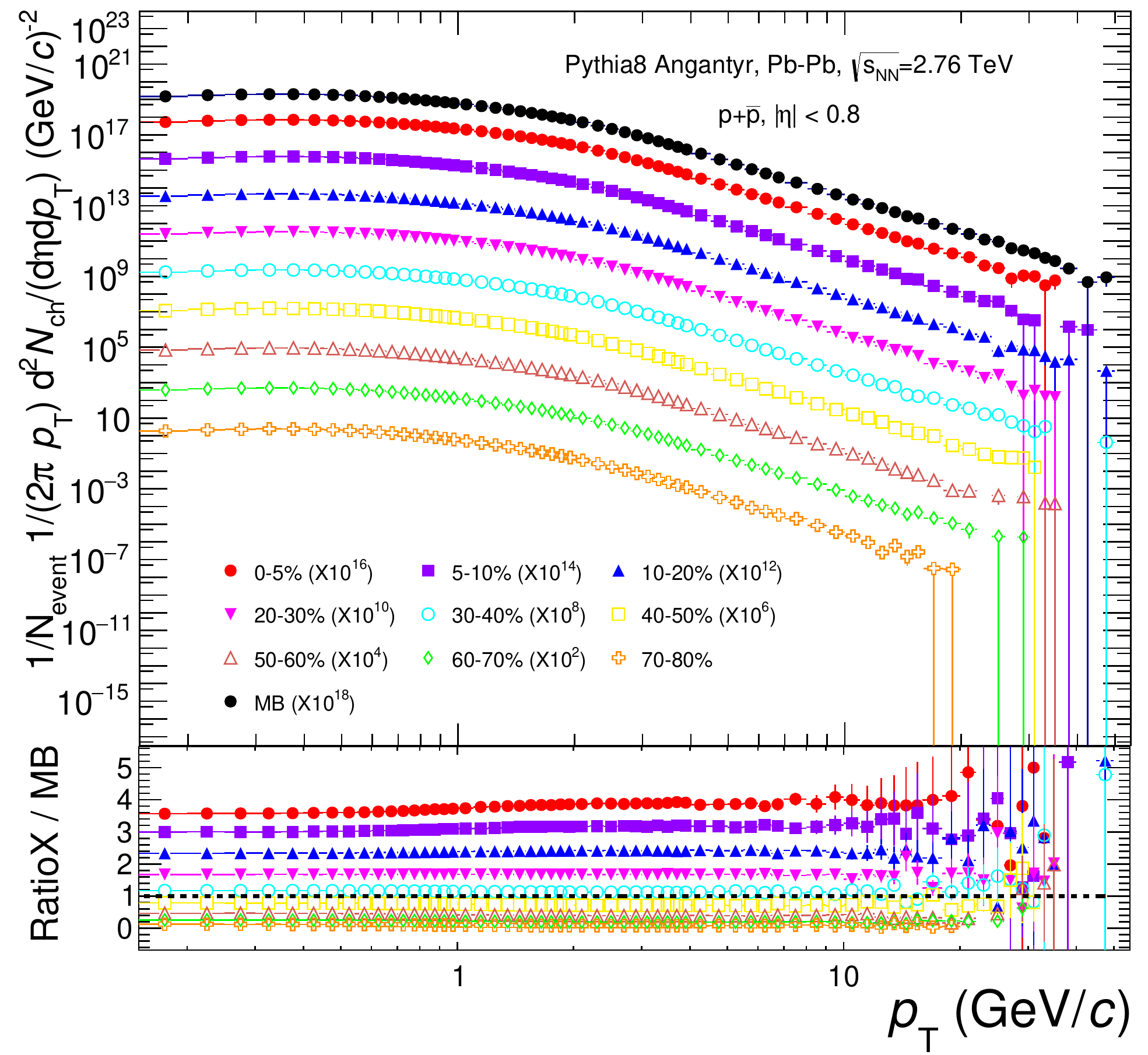}
\caption{(Color Online) \pt \ spectra of identified charged-particles ($\pi^{\pm}$,K$^\pm$, p$(\bar{\rm p})$) in various centrality classes. The lower panel show the ratios for each centrality class to MB.}
\label{fig3} 
\end{figure*}

FIG.~\ref{fig3}. shows the \pt \ spectra of identified charge-particles $\pi^{\pm}$,$K^{\pm}$, and $p(\bar{p})$ in different centrality classes and for minimum bias (MB). The spectra were obtained using the same selection cuts in all charged-particles species. To visualize better, we multiplied scale factors to each \pt \ spectra. From the lower panel of FIG.~\ref{fig3}., the \pt \ spectra corresponding to (30-40)\% centrality is seen to coincide with the MB spectra. We report a shift in the hardness of the \pt \ spectra for each centrality class with respect to MB. For the threshold centrality class (30-40)\%, classes (0-5)\%, (5-10)\%, (10-20)\%, (20-30)\% are harder while classes (40-50)\%, (50-60)\%, (60-70)\% and (70-80)\% are softer with respect to MB. It is to be noted that a similar trend is observed for all the identified particles.

\begin{figure*}[ht!]
\includegraphics[scale=0.37]{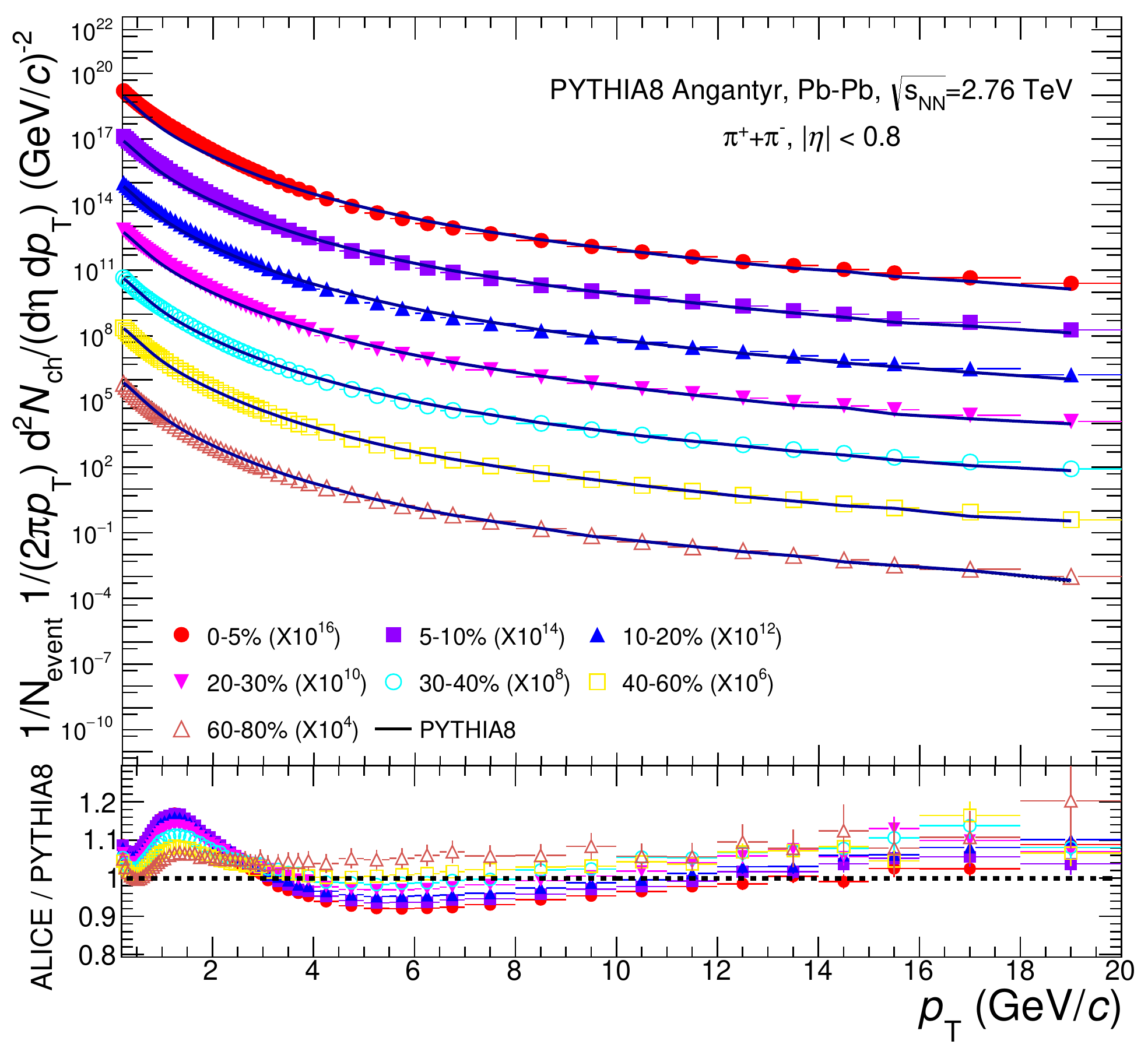}
\includegraphics[scale=0.37]{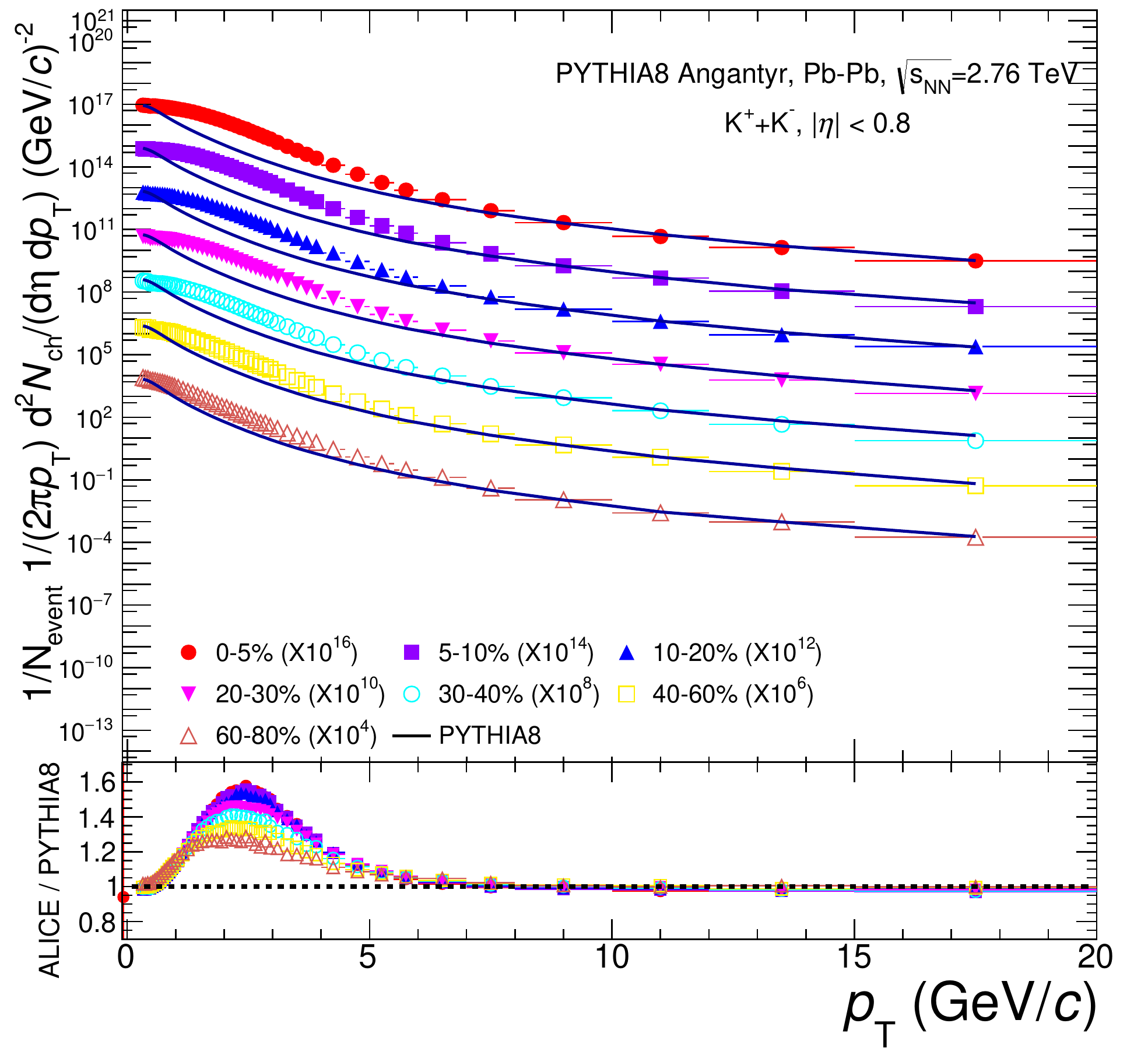}
\includegraphics[scale=0.37]{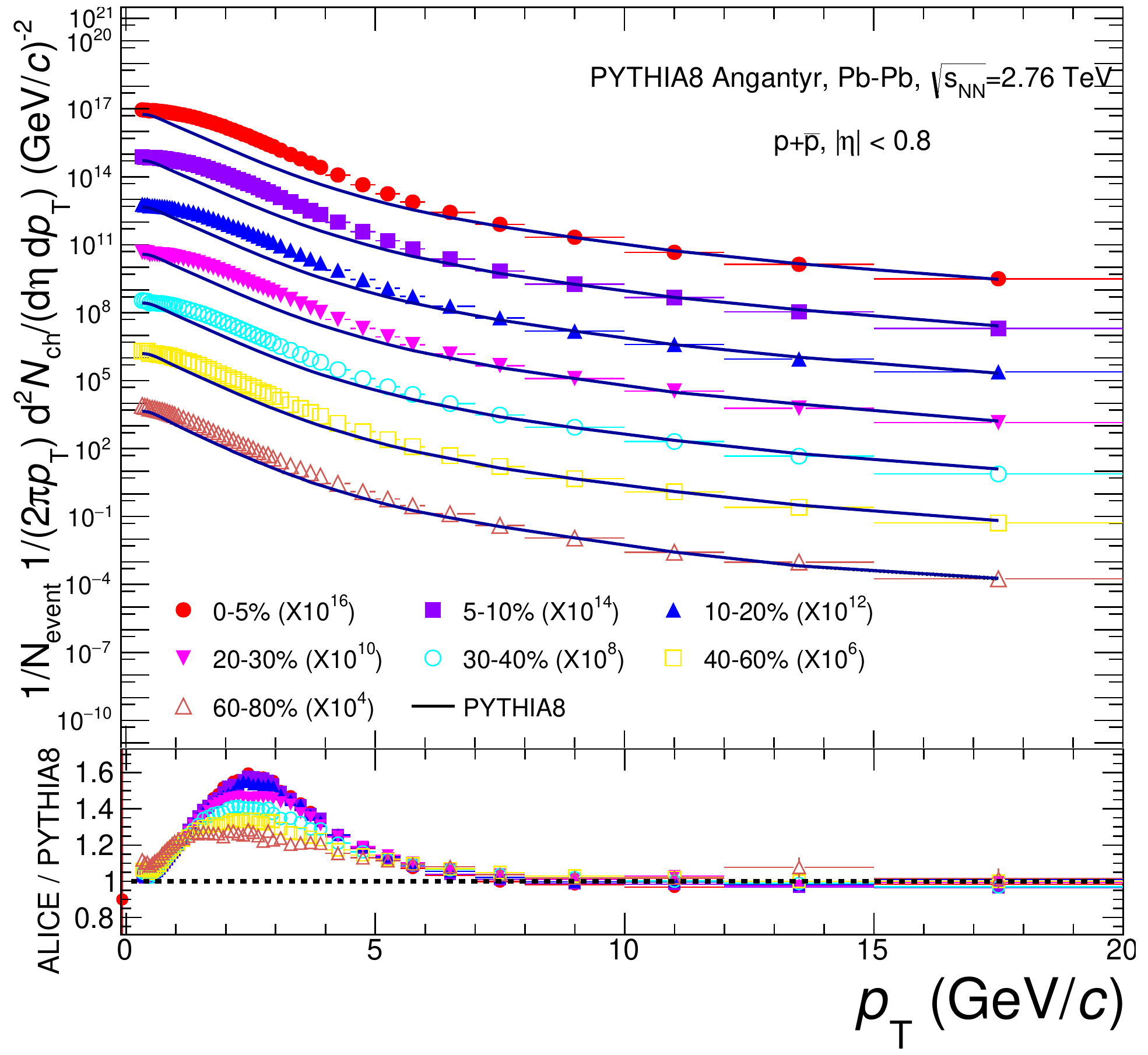}
\caption{(Color Online) \pt \ spectra of identified charge particles ($\pi^{\pm}$,K$^\pm$, p$(\bar{\rm p})$)  compared to ALICE measurements. The lower panel show the ratios for each centrality class to data.}
\label{fig3data} 
\end{figure*}

\begin{figure*}[ht!]
\includegraphics[scale=0.37]{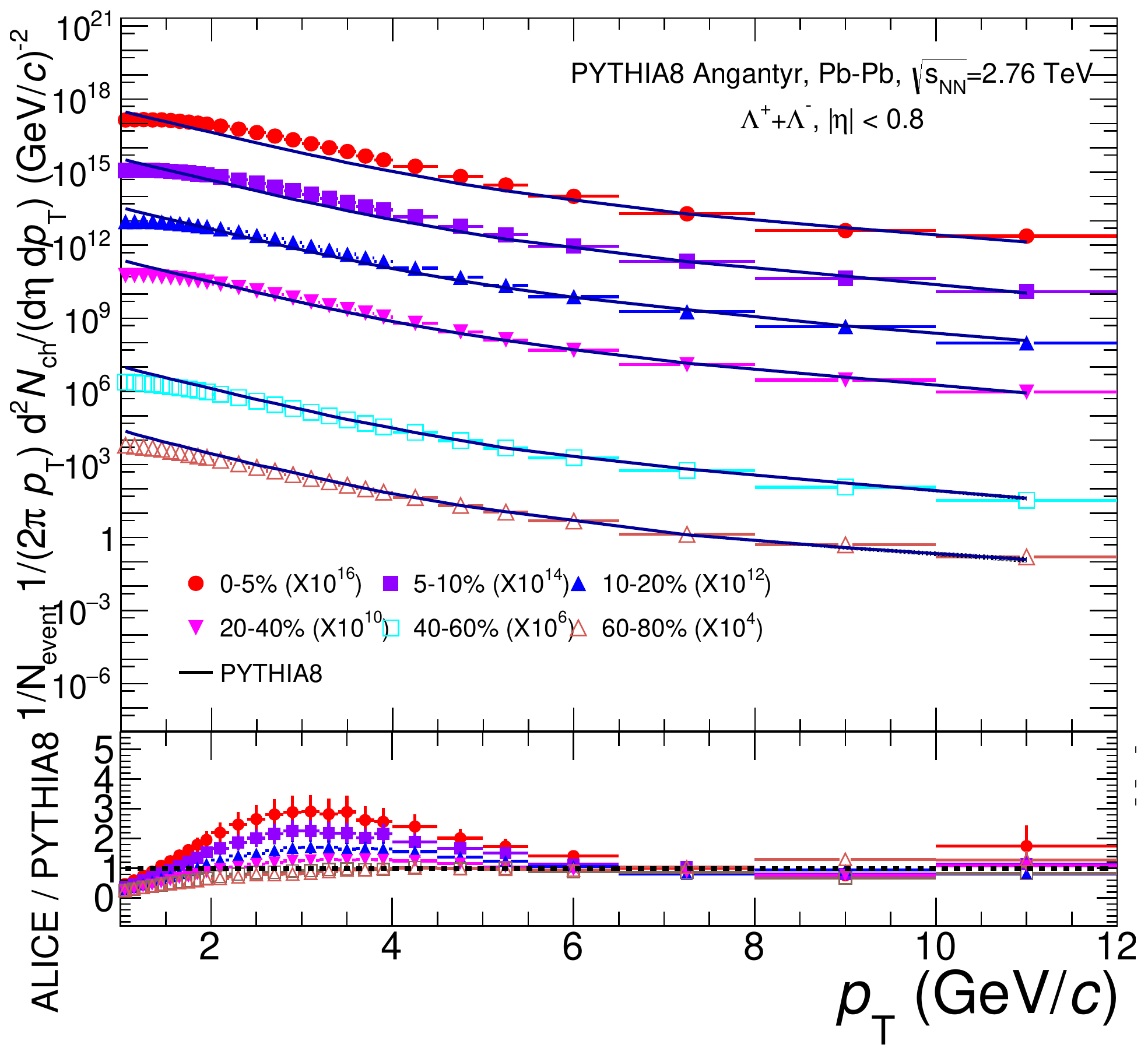}
\includegraphics[scale=0.37]{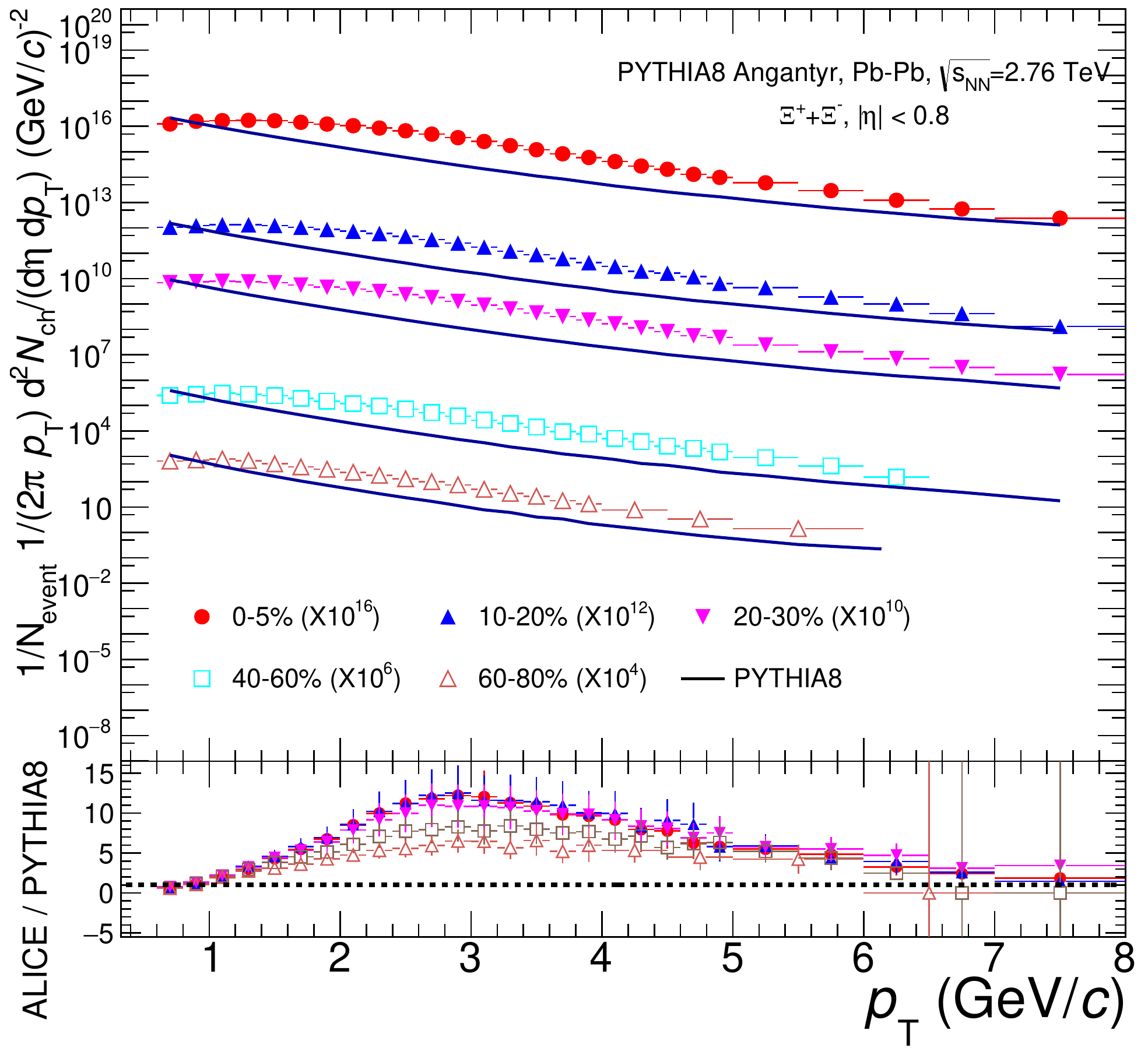}
\includegraphics[scale=0.37]{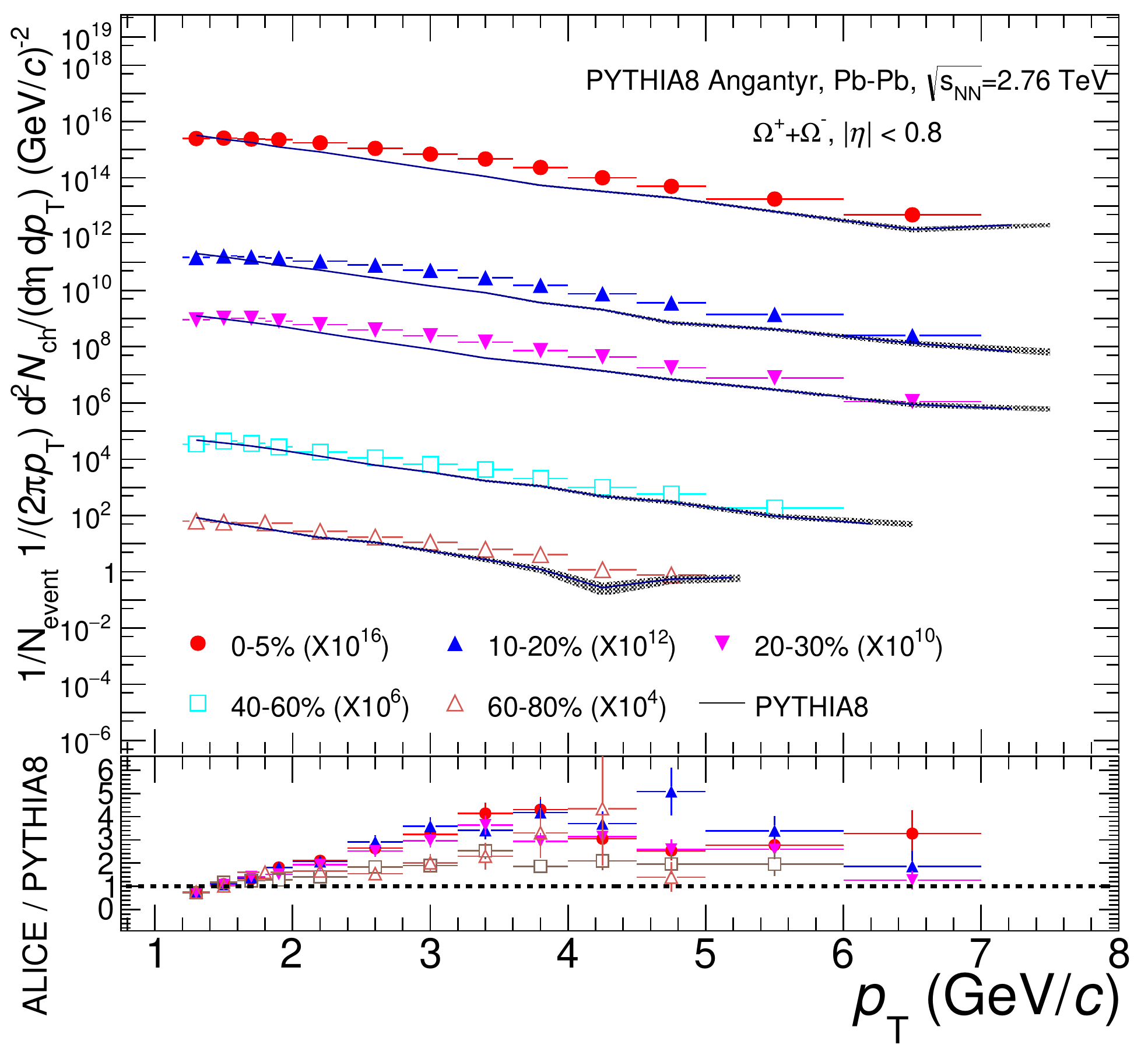}
\caption{(Color Online) \pt \  spectra of strange and multi-strange baryons ($\Lambda^{\pm}$, $\Xi^{\pm}$, $\Omega^{\pm}$). The lower panel show the ratios for different centrality classes to data.}
\label{fig3lxi} 
\end{figure*}

\begin{figure*}[ht!]

\includegraphics[scale=0.41]{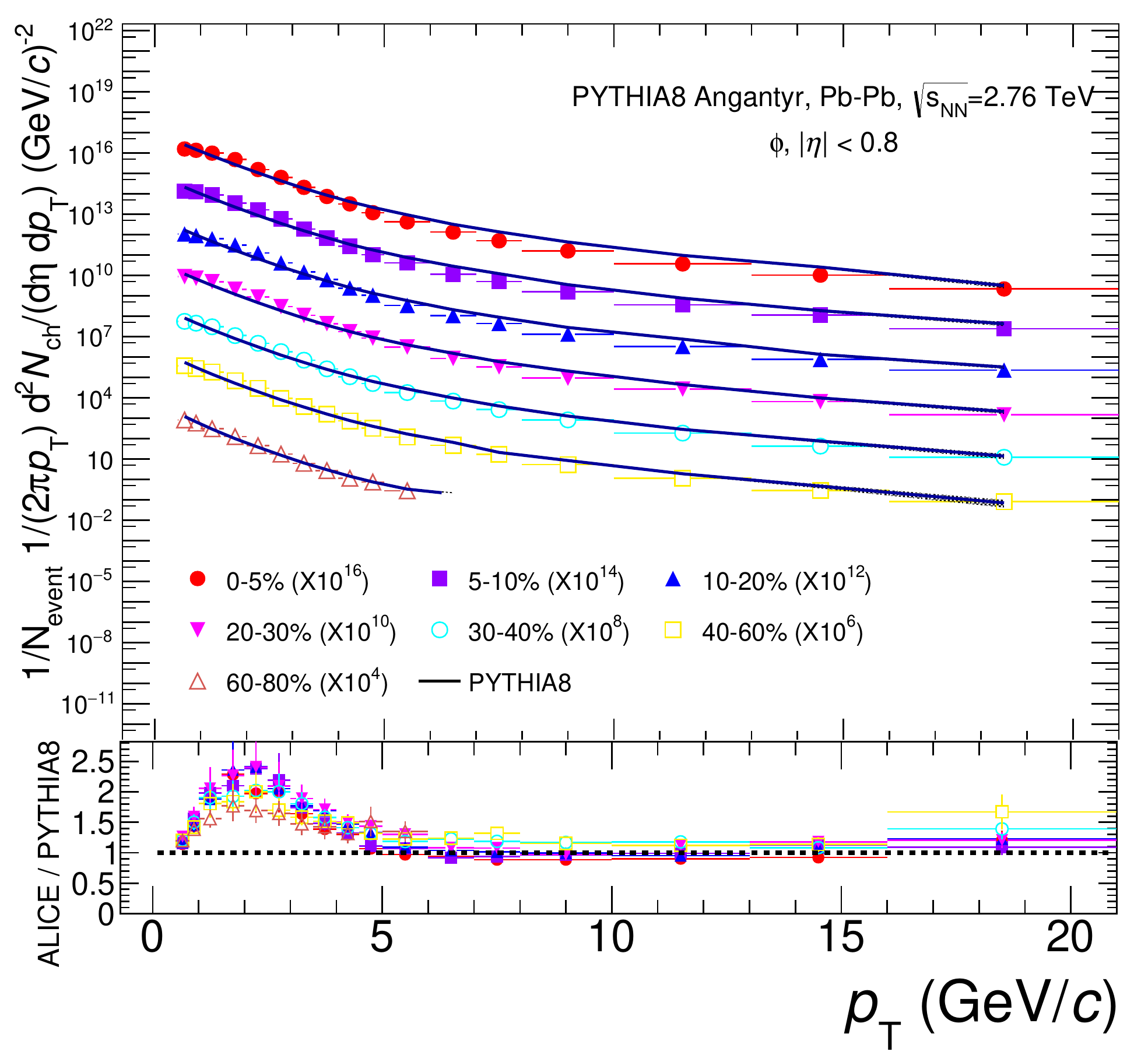}
\includegraphics[scale=0.40]{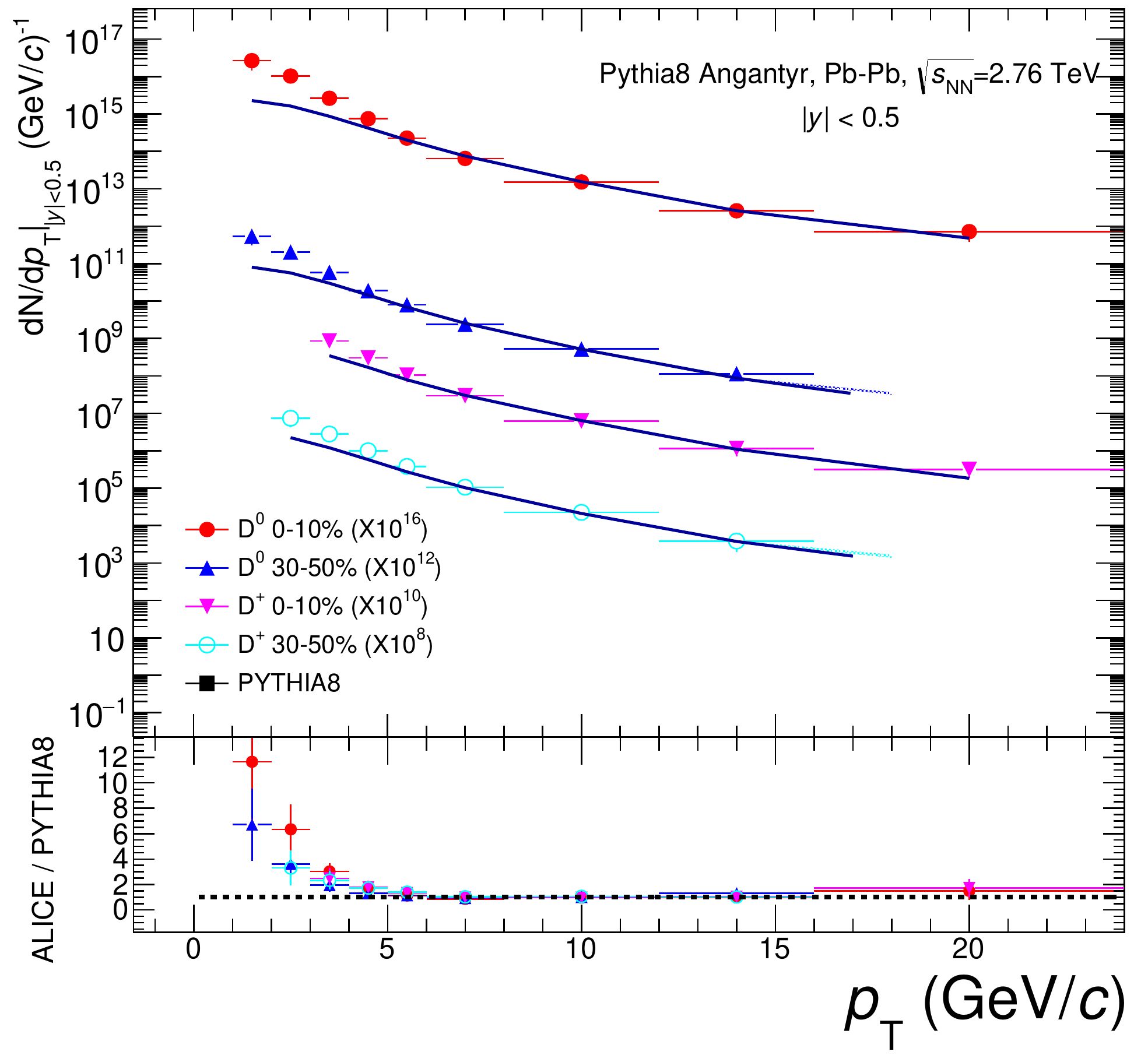}
\caption{(Color Online) \pt \  spectra of $\phi$ and D-mesons. The lower panel show the ratios for each centrality class to data.}
\label{fig3ophi} 
\end{figure*}

The \pt \ spectra of hadrons obtained at the final state are compared to measurements from the ALICE Collaboration. In FIG.~\ref{fig3data}. the PYTHIA predictions find a good match with the data for pions. For protons and kaons, we report a slight underestimation by PYTHIA at low \pt . The hump at low \pt \ is probably due to the pQCD implementation of PYTHIA, whereas we expect NRQCD effects in this regime. We also compare strange baryon \pt \ spectra ($\Lambda$, $\Xi$, $\Omega$) with ALICE in all centrality ranges considered, as shown in FIG.~\ref{fig3lxi}. The ratios show a similar peak at \pt $\sim$ 3 GeV/$c$. It is also observed that the width of the hump increases with strangeness and mass, especially for central and semi-central events. At higher centralities, the strange baryons show good compatibility. In FIG.~\ref{fig3ophi}., we compare the $\phi$ meson and D-meson (D$^0$, D$^+$) \pt \ spectra, where the $\phi$ meson spectra are seen to be consistent with ALICE measurements in all centralities. In the case of D-mesons, we see PYTHIA predictions depart at low \pt , however in good agreement at intermediate-higher values. This helps us conclude that the more prominent peaks observed for strange baryons (FIG.~\ref{fig3lxi}.) have a strangeness dependence rather than mass.


\begin{figure*}[ht!]
\includegraphics[scale=0.40]{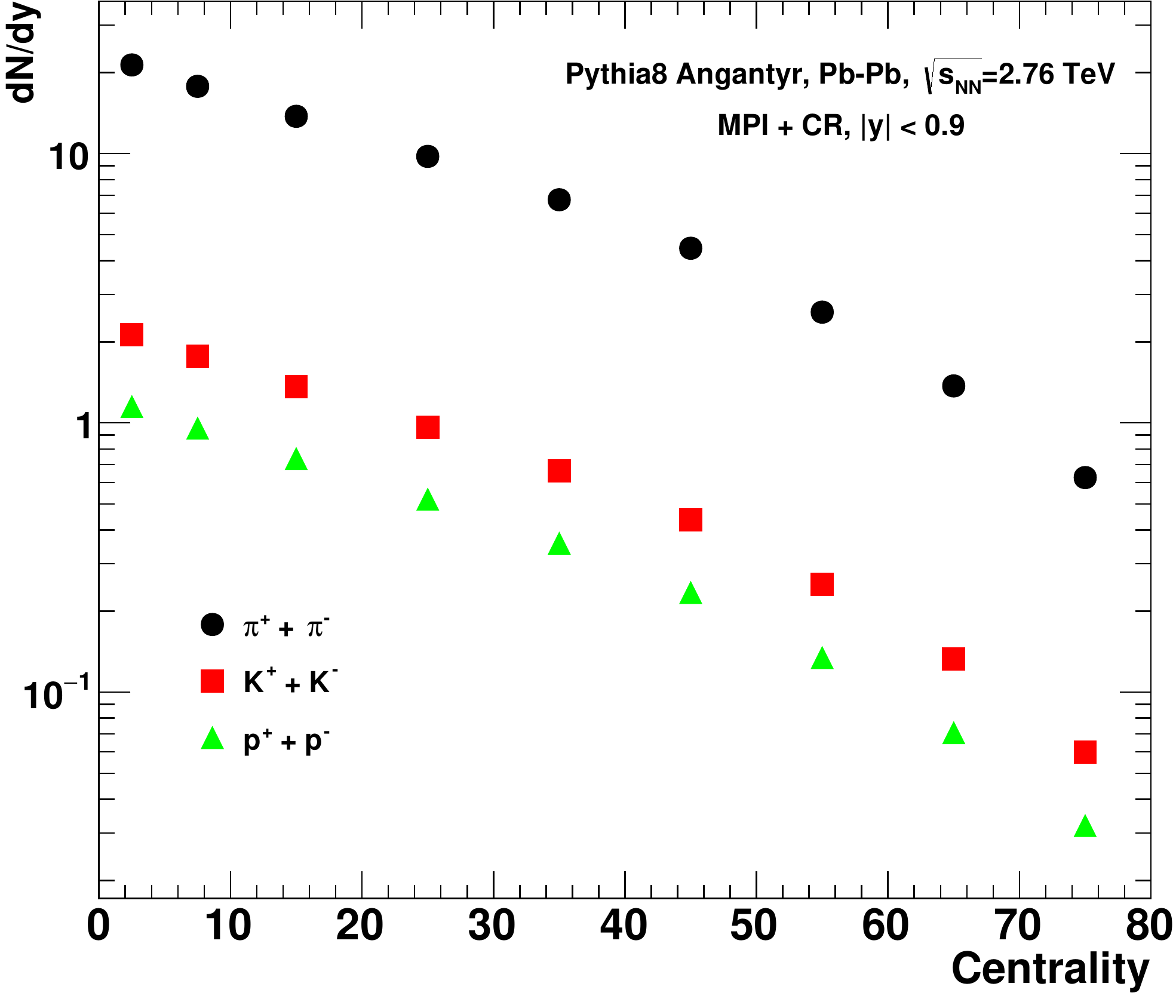}
\includegraphics[scale=0.40]{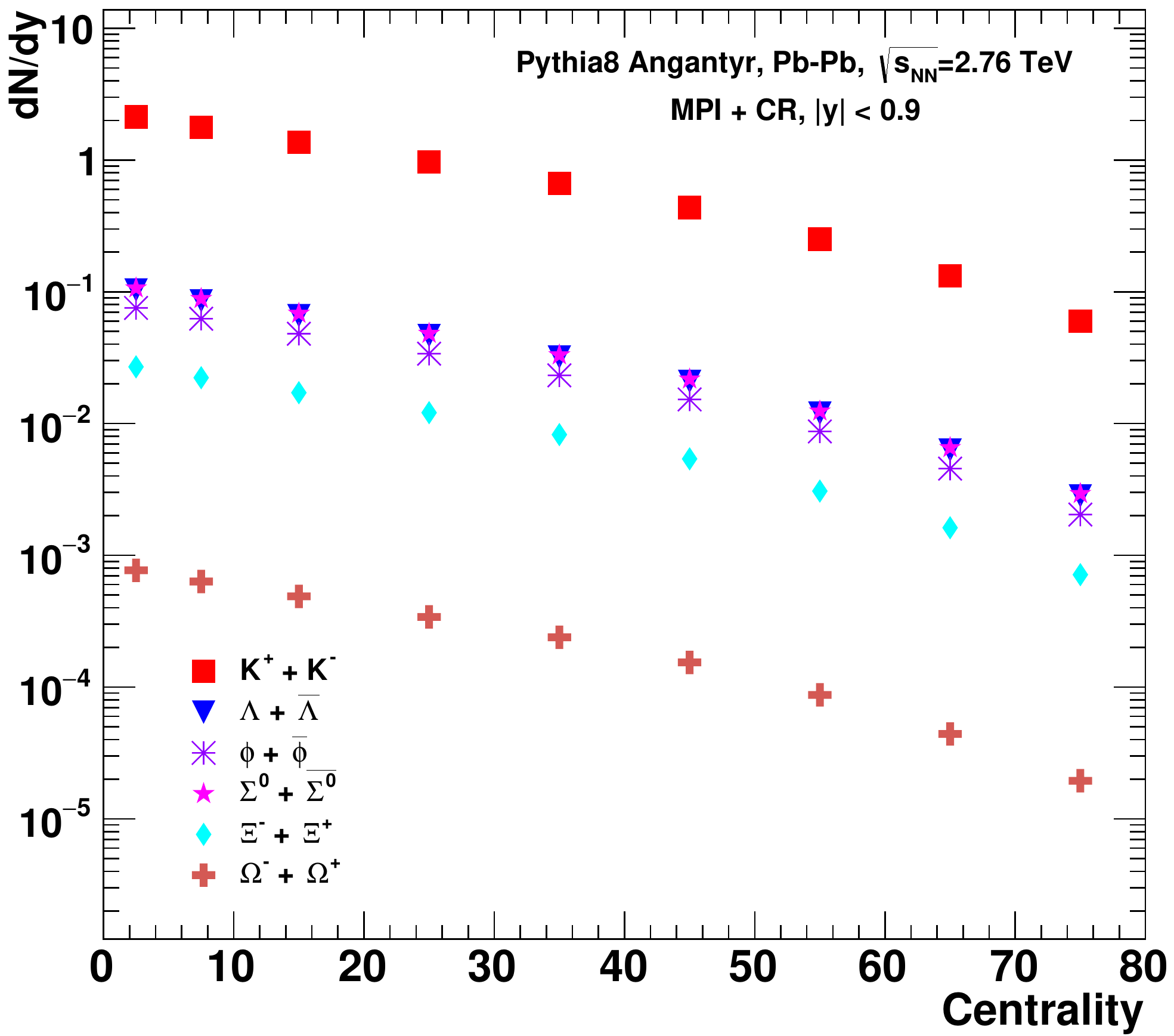}
\caption{(Color Online) Yield of identified particles (Left) and strange particles (Right) as a function of centrality.}
\label{fig4}
\end{figure*}

\subsection{\pt \  integrated yield of identified and strange particles}
 \label{mult_CSB_C_s_2}
 
The \pt \ integrated yields of $\pi^{\pm}$,$K^{\pm}$ and $p(\bar{p})$ are shown in FIG.~\ref{fig4} (Left) within rapidity range -0.9 $<y<$ 0.9 normalized by the total number of events. It is observed that the yields of different particles are increasing going from peripheral to most central. We can see a clear mass ordering in yields, with lower mass pions having higher yields, while protons being heavier have lower yields. This is expected towards central collisions; the probability for hard scatterings will be higher, resulting in high particle production. Production of a lighter particle requires lesser energy as compared to a heavier particle and will be more dominant in peripheral collisions. This gives further concreteness towards the use of PYTHIA8 Angantyr with the current configuration for heavy-ion collisions.

In FIG.~\ref{fig4} (Right), \pt -integrated yields of strange particles are shown. One can see the same features in strange particles like that of identified particles observed in FIG.~\ref{fig4} (Left). Production of strange particles is seen to reduce towards peripheral collisions with a similar trend in mass, except for $\rm \phi$ mesons.

\subsection{Particle ratios}

\begin{figure*}[t!]
\includegraphics[scale=0.42]{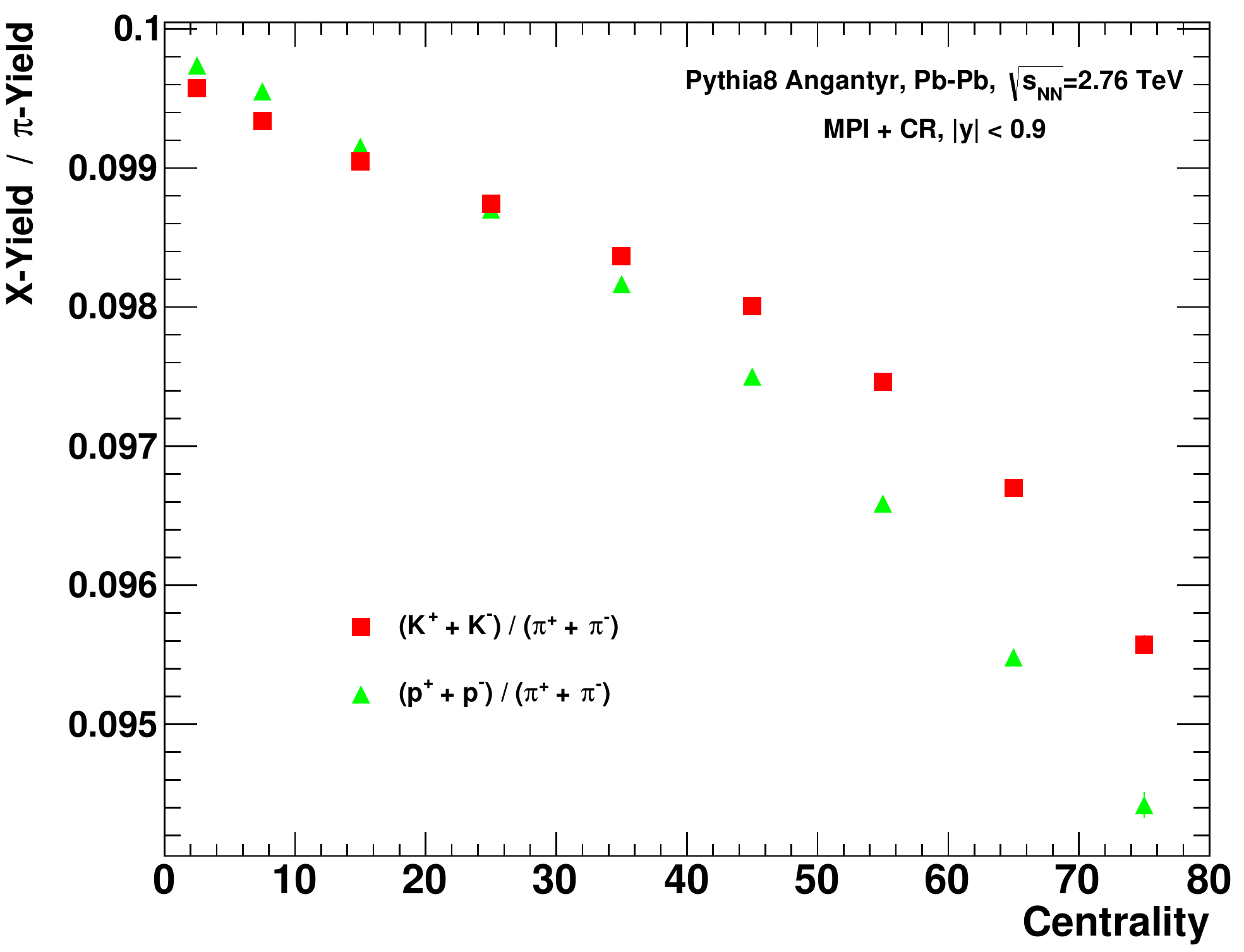}
\includegraphics[scale=0.39]{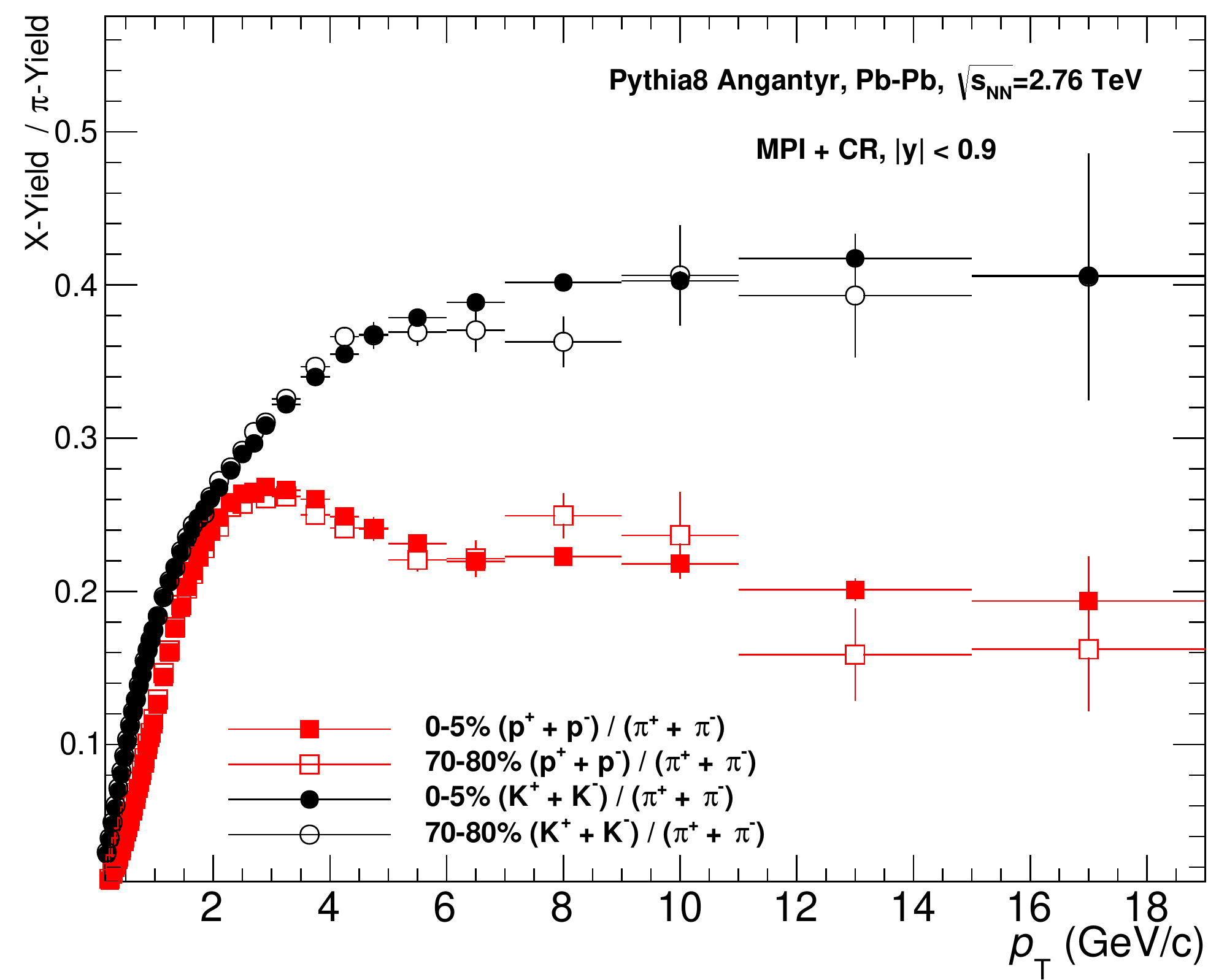}
\caption{(Color Online) Ratio of yields of identified particles over $\pi^+ + \pi^-$ as a function of centrality (Left) and as a function of transverse momentum (Right).}
\label{fig6} 
\end{figure*}

By the bare yield distribution, we cannot quantitatively measure the enhancement or suppression of different particle species. The best way to do this is to estimate the yield with respect to other particles. We measure the ratio of proton and kaon yields over pions to inspect the variation over centrality and \pt . FIG.~\ref{fig6}. shows the measured yield ratios vs. centrality (Left) and \pt  (Right). We scale proton over pion ratios for every centrality class for better comparison with the corresponding quantity versus \pt \ (Left). The scale factor is calculated using the formula:

$$Scale \ factor = \dfrac{K/\pi}{X/\pi}$$

Here X refers to different particle species. We can see from this FIG.~\ref{fig6}. that both ratios are increasing towards most central; however proton over pion ($p/\pi$) ratio drops rapidly than kaon over pion ($K/\pi$). $K/\pi$ ratio increases at lower \pt \ but decreases at higher \pt , showing a bump around 3 GeV/c. In heavy-ion collisions, this is the consequence of radial flow, but in PYTHIA, this is attributed to color reconnection. We can argue that CR could be another mechanism of flow where a longitudinal boost is implemented at the initial state (partonic state), just before hadronization starts. Understanding this mechanism is important, as it could provide an explanation of flow-like patterns. At higher \pt , more particles correspond to jets, in which these particles become insensitive to CR. If one increases MPI, we see an enhancement in the bump region.  We do not observe any bump in the $K/\pi$ ratio. It increases at lower \pt \ and then turns flat at higher \pt \ values. Here, we use the default tune of CR that is implemented in PYTHIA. By tuning CR, studies report this effect can also be observed in meson to meson ratio~\cite{OrtizVelasquez:2013ofg}. 

\begin{figure*}[ht!]
\includegraphics[scale=0.41]{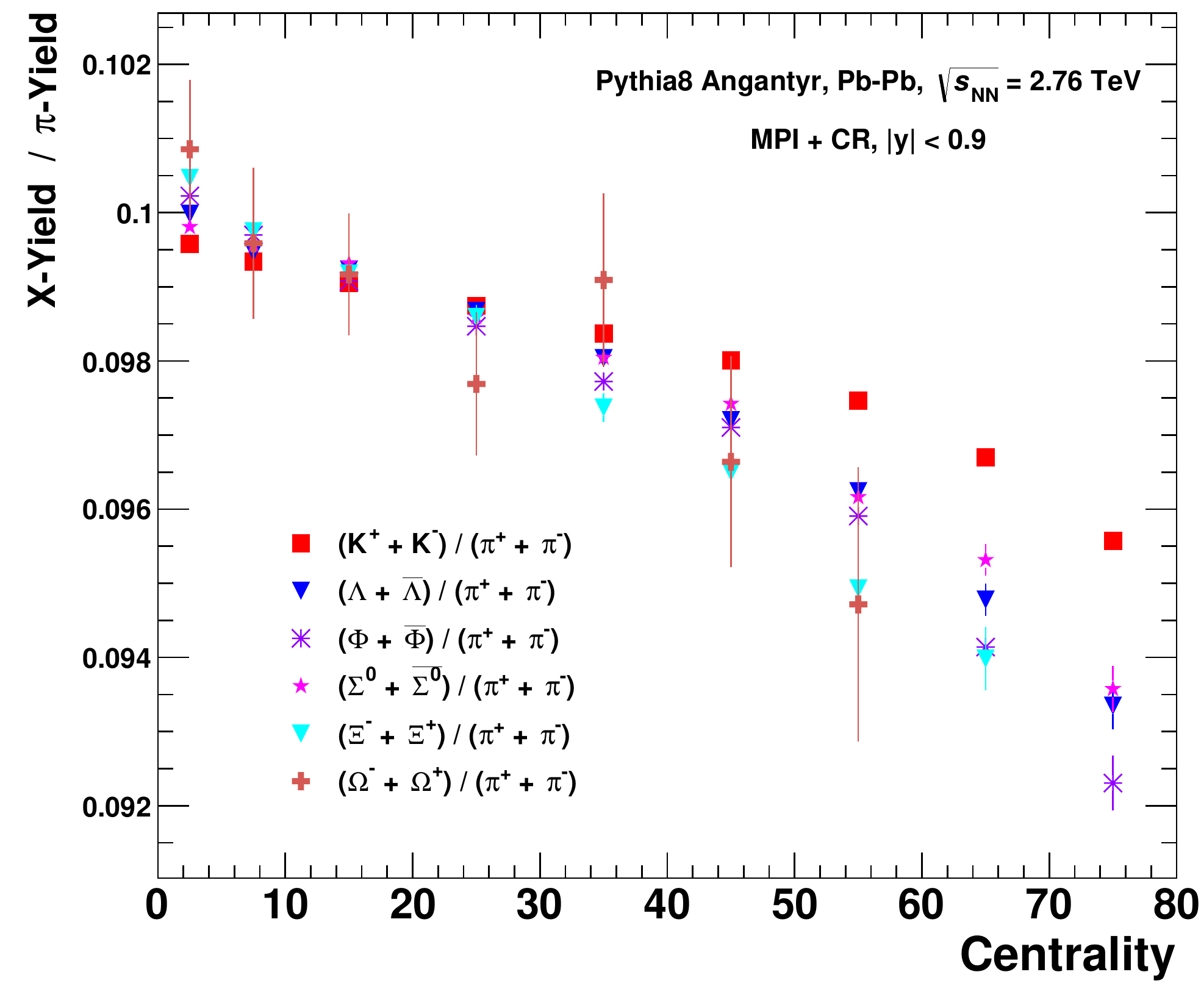}
\includegraphics[scale=0.38]{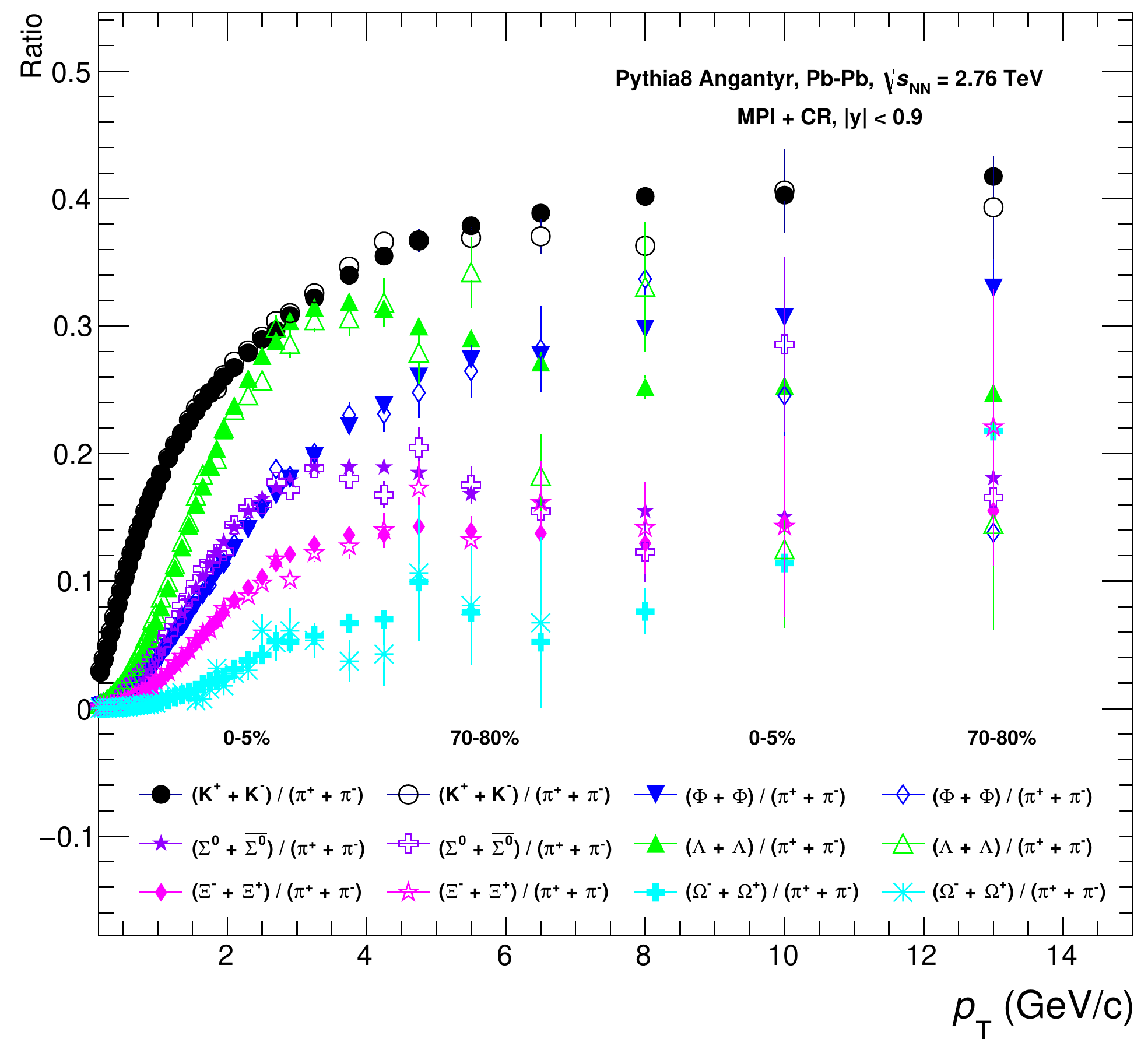}
\caption{(Color Online) Ratio of yields of strange particles over $\pi^+ + \pi^-$ as a function of centrality (Left) and as a function of transverse momentum (Center).}
\label{fig7} 
\end{figure*}

Similarly, we show yield ratios of strange particles over pions as a function of centrality in FIG.~\ref{fig7} (Left). Each yield ratio of different particles is scaled with a similar method as mentioned before. We observe a clear strangeness enhancement as we go from most central to peripheral collisions. Heavy strange particle ratios are showing more enhancement, as reported in FIG.~\ref{fig7} (Left), slopes of strange particle ratios increase towards heavier strange particles. This is due to overlapping color strings forming (ropes) at higher densities~\cite{Bierlich:2018xfw}. In FIG.~\ref{fig7}, we show yield ratios of strange particles as a function of transverse momentum in two different centrality classes at 0-5\% and 70-80\%. For all strange particles, yield ratios increase towards higher \pt . As expected, the ratio of yields is lesser for strange heavy particles. We also conclude by observing FIG.~\ref{fig6} (Right) and FIG.~\ref{fig7} (Right) that meson to pion ratios are not showing the bump but shows for baryon to pion ratios.

\begin{figure*}[ht!]
\includegraphics[scale = 0.4]{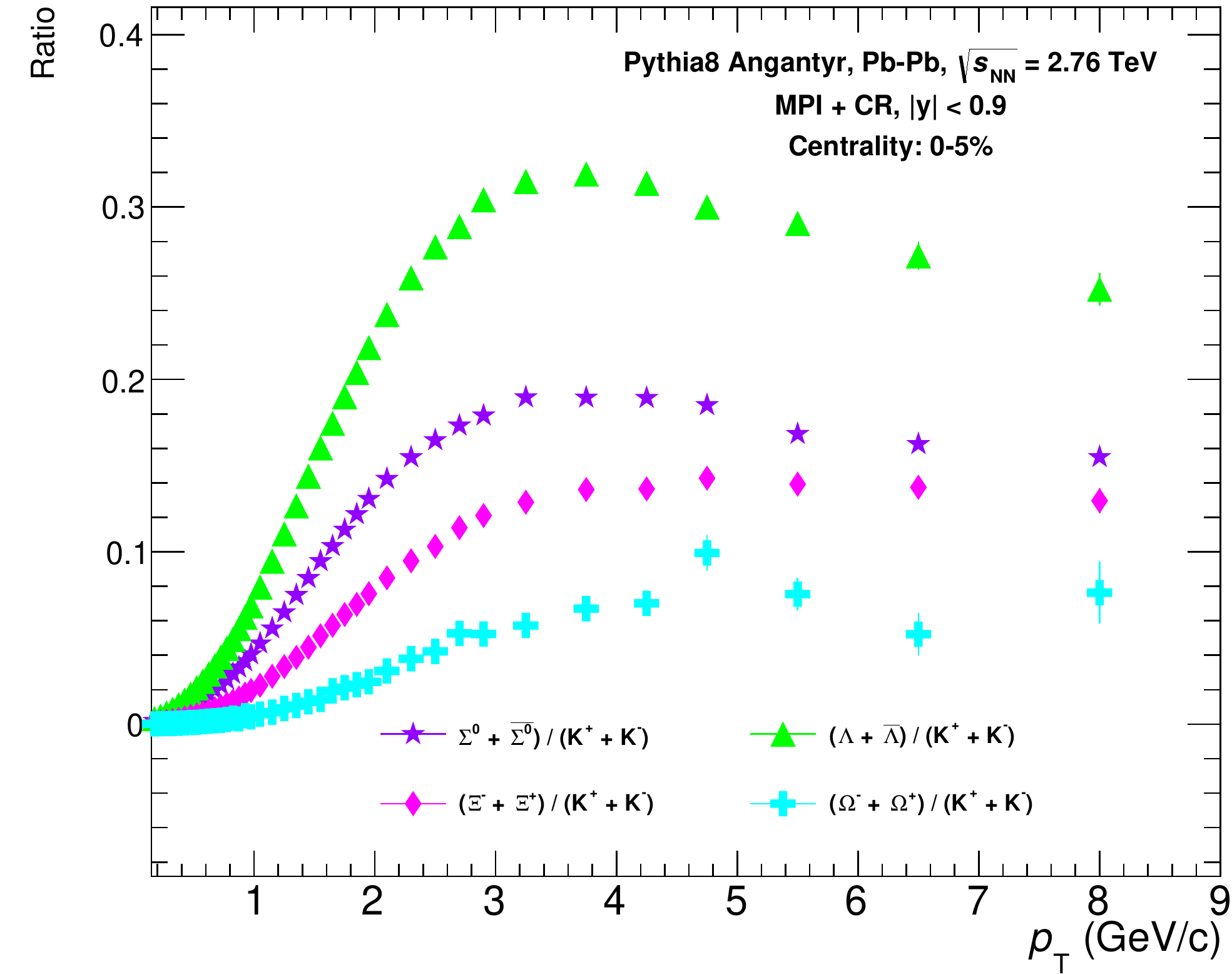}
\caption{(Color Online) Ratio of yields of strange particles over ($ K^{+}+ K^{-}$) as a function of transverse momentum.}
\label{fig8} 
\end{figure*}

In FIG.~\ref{fig8}. we show the ratio of strange particles over ($K^{+}+K^{-}$) mesons. The ratio increases as \pt \  increases, and after a peak close to 3-4 GeV/c, it decreases. The position of the peak shift towards higher \pt \ for strange heavy particles. A study reports a similar type of observation seen in experimental data for Pb--Pb collisions~\cite{ALICE:2013cdo}. This effect is generally seen in heavy-ion collisions as a consequence of radial flow~\cite{Fries:2003fr}.

 \section{Summary}
  \label{sum}
 In this work, we attempt to study the dynamics of particle production in Pb--Pb collisions at $\sqrt{s_{\rm NN}}$ = 2.76 TeV using the Angantyr model incorporated in PYTHIA8. This model is an extrapolation of pp collision into pA and AA collisions without a thermalized medium. Without any collectivity present, we assume that MPI with CR mimics flow-like features~\cite{Maldonado-Cervantes:2014tva}. In this regard, the primary observations of this work are summarised below:
  
  \begin{itemize}
  
  \item[$\bullet$] The multiplicity distribution of charged particles with different combinations of MPI and CR tunes is compared to ALICE measurements. It is observed that setting CR off, PYTHIA8 Angantyr predictions overestimate ALICE data, whereas turning MPI off underestimates the data. MPI with CR setting is observed to be a good tune to match the multiplicity distribution obtained from PYTHIA with ALICE results.
  
  \item[$\bullet$] The \pt \ spectra simulated by PYTHIA8  Angantyr at different centralities are also consistent with experimental data within the uncertainties. It is also observed that the $\langle$\pt$\rangle$ distribution obtained by using MPI + CR is consistent with ALICE data.
   
  \item[$\bullet$] \pt \ spectra of identified particles ($\pi^{\pm}$,$K^{\pm}$,$p(\bar{p})$) in different centrality classes along with minimum bias (MB) is obtained. Hardening of the \pt \ spectra can be seen towards most central and is more pronounced for heavier particles. A centrality cut-off at (30-40)\% is observed and found to coincide with MB, which implies centralities above this cut-off produce harder particles.

 \item[$\bullet$] The \pt \ integrated yield normalized by total events of identified charged particles and strange particles show a clear mass ordering. Heavier particle production is lesser and decreases towards peripheral collisions. The rate of production is higher for protons compared to kaons towards central collisions to pions.
  
  \item[$\bullet$] It is observed that the ratio of kaon over pion as a function of \pt \ increases with increasing \pt \ which then saturates toward higher \pt. In contrast, the ratio of proton over pion shows a peak around 3 GeV/c due to CR with MPI in PYTHIA8, which mimic flow-like patterns. A similar rise is observed for all the strange baryon over pion ratios, although unobserved for the meson to meson ratio.

  \item[$\bullet$] It is observed that the slope of strange particles to pion ratio as a function of centrality is more significant for strange heavy particles. We report that the peak of strange baryon to pion ratio and strange baryon to kaon as a function of \pt \  shifts toward higher \pt \ for heavier strange particles. This shows that strangeness enhancement is dominant in strange heavy particles, which is a consequence of color strings overlapping at higher densities in accordance with the CR mechanism.  
  
   \end{itemize}
   
\clearpage

From the present study and the subsequent observations, it can be concluded that MPI with CR mimics flow-like features, and one can use this tune in PYTHIA8 Angantyr to carry out heavy-ion studies. Although we have shown features resembling radial flow, one can also observe higher-order flow harmonics ($v_2$,$v_3$) in models that pertain to QGP free collision systems.~\cite{daSilva:2020cyn}. 

\section{Acknowledgement} 
The author acknowledges the financial support by the DST-INSPIRE program of the Government of India.


 \end{document}